# NarrowBand IoT Data Transmission Procedures for Massive Machine Type Communications

Pilar Andres-Maldonado, Pablo Ameigeiras, Jonathan Prados-Garzon, Jorge Navarro-Ortiz, and Juan M. Lopez-Soler

*Department of Signal Theory, Telematics, and Communications*

*University of Granada*

*Abstract-* Large-scale deployments of massive Machine Type Communications (mMTC) involve several challenges on cellular networks. To address the challenges of mMTC, or more generally, Internet of Things (IoT), the 3rd Generation Partnership Project has developed NarrowBand IoT (NB-IoT) as part of Release 13. NB-IoT is designed to provide better indoor coverage, support of a massive number of low-throughput devices, with relaxed delay requirements, and lower-energy consumption. NB-IoT reuses Long Term Evolution functionality with simplifications and optimizations. Particularly for small data transmissions, NB-IoT specifies two procedures to reduce the required signaling: one of them based on the Control Plane (CP), and the other on the User Plane (UP). In this work, we provide an overview of these procedures as well as an evaluation of their performance. The results of the energy consumption show both optimizations achieve a battery lifetime extension of more than 2 years for a large range in the considered cases, and up to 8 years for CP with good coverage. In terms of cell capacity relative to SR, CP achieves gains from 26% to 224%, and UP ranges from 36% to 165%. The comparison of CP and UP optimizations yields similar results, except for some specific configurations.

## 1. Introduction

The successful deployment of massive Machine Type Communications (mMTC) will be a key aspect for a large number of heterogeneous verticals, ranging from smart cities, e-health to industrial IoT and much more. Recently, Mobile Networks (MNs) have been considered a convenient option to provide connectivity to Internet of Things (IoT) devices. Particularly, MNs enable ubiquitous coverage, mobility and facilitate interworking with short-range wireless networks.

MNs have been designed for high-performance mobile broadband communications. However, mMTC has very different characteristics as it typically involves the automatic sending of infrequent and non-delay sensitive low volume data by a massive number of devices. Hence, to support mMTC, MNs face several major challenges [1]:

- Ultra-low device complexity. For mass deployment, mMTC devices need to be cheap. Therefore, they may offer very limited performance.
- High network scalability. The network has to support a huge number of low throughput devices. Forecasts indicate that the number of connected mMTC devices could range a factor 10x to 100x more devices than mobile phones.
- Efficiently accommodate small burst of data.
- Long battery lifetime. The energy consumption should be reduced as many devices will be battery-powered, and often the cost of replacing batteries in the field is not viable. The goal is to allow battery lifetime of more than 10 years with a battery capacity of 5 Wh.
- High Coverage. Improved indoor coverage of 20 dB compared to legacy General Packet Radio Service (GPRS), corresponding to a Maximum Coupling Loss (MCL) of 164 dB.

To address these challenges in Cellular IoT (CIoT), the 3rd Generation Partnership Project (3GPP) has developed the NarrowBand IoT (NB-IoT) concept as part of Release 13. NB-IoT is designed to provide better indoor coverage, support of a massive number of low-throughput devices, with relaxed delay requirements and lower-energy consumption [1]. NB-IoT reuses Long Term Evolution (LTE) functionality with simplifications and optimizations. Particularly for small data transmissions, NB-IoT specifies two procedures to reduce the required signaling [2]:

i) Control Plane CIoT Evolved Packet System (EPS) optimization (CP)

ii) User Plane CIoT EPS optimization (UP).

These procedures improve the accommodation of mMTC's small bursts of data, compared to the conventional procedures of LTE mobile networks. The current design of LTE requires that the User Equipment (UE) performs the connection establishment (through a Service Request (SR) procedure), before sending the data packets. After completing the data transmission, the connection is released (by means of a S1 Release procedure). Hence, the efficiency of this scheme is low for handling small amounts of data.

In this work, we provide an overview and an analysis of the novel data transmission procedures specifically designed for mMTC in NB-IoT. We compare them with the SR procedure. More precisely, we evaluate two main performance indicators of NB-IoT: i) energy consumption and ii) radio resource consumption. The presented results summarize four transmission scenarios and three coverage levels. For each of them, we study the main factors affecting the performance indicators of all procedures. The results show both optimizations obtain better results compared to the SR procedure. Additionally, the comparison of CP and UP optimizations yields similar results, except for some specific configurations.

The paper is organized as follows. Section 2 summarizes NB-IoT technology. Section 3 describes CP and UP schemes and describes their differences with SR. Section 4 lists some new functionalities required in LTE by these optimizations. Then, Section 5 presents the results of our evaluation. Finally, Section 6 draws the main conclusions.

## 2. NB-IoT

NB-IoT focusses on low-cost devices with lower-energy consumption and higher coverage requirements. NB-IoT meets these key demands by means of i) the use of a small portion of the existing available spectrum, ii) a new radio interface design, iii) simplified LTE network functions.

### *2.1 NB-IoT radio design*

The new NB-IoT radio interface design is derived from the legacy LTE. The NB-IoT carrier has a 180 kHz bandwidth with support for multi-carrier operation. In downlink, Orthogonal Frequency-Division Multiple Access (OFDMA) is applied using a 15 kHz subcarrier spacing over 12 subcarriers with 14 symbols used to span a subframe of 1 ms. In uplink, Single Carrier Frequency Division Multiple Access (SC-FDMA) is applied, using either 3.75 kHz or 15 kHz subcarrier spacing.

NB-IoT physical channels and signals are primarily multiplexed in time. Figure 1 shows an example of NB-IoT subframes design. For more information on NB-IoT, see [3].

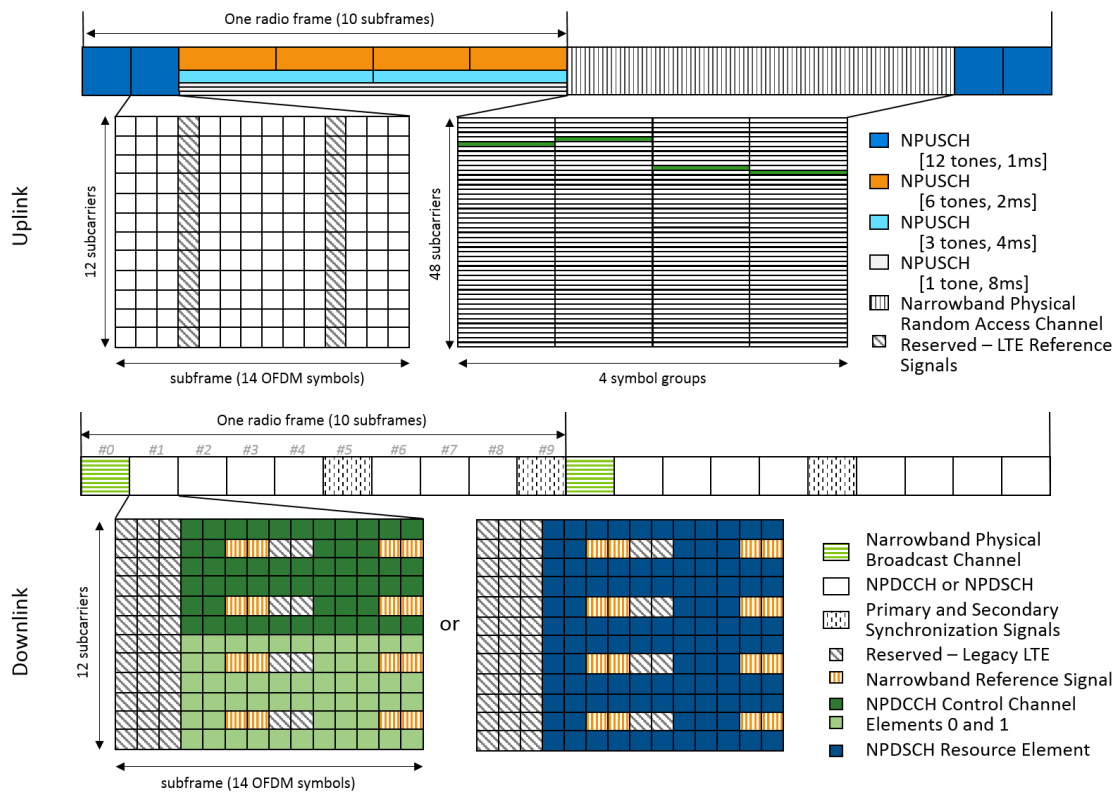

Figure 1: NB-IoT physical channels time multiplexing [3].

NB-IoT defines three operation modes to provide deployment flexibility:

- Stand-alone: utilizing, for example, one or more GSM carriers.
- Guard-band: utilizing the unused resource blocks within an LTE carrier's guard-band.
- In-band: utilizing resource blocks within an LTE carrier.

Additionally, NB-IoT uses the concept of repetitions and signal combining techniques to improve coverage extension [4]. To serve UEs in different coverage conditions that have different ranges of path loss, there may be up to three coverage enhancement (CE) configurations in the random access with their specific settings. After that, the eNodeB selects the configuration of the radio resources, the Modulation and Coding Scheme (MCS) and repetitions depending on UE's coverage.

*2.2 Energy efficiency*

NB-IoT is designed for long-life devices and targets a battery lifetime of more than 10 years. To this end, NB-IoT reuses LTE's power saving mechanisms but extending the timers involved to achieve longer battery lifetime. In LTE, there are two key power saving mechanisms: Discontinuous Reception (DRX) and Power Saving Mode (PSM). Both mechanisms modify the way the UE communicates with the network. This communication requires a Radio Resource Connection (RRC) established between the UE and the eNB. There are two possible RRC connection states: Connected and Idle. A UE in RRC Connected state has an active RRC connection. Therefore, the eNB can directly allocate resources to the UE. Otherwise, a UE is in RRC Idle state. The UE can transit from RRC Connected to RRC Idle state because of different causes, such as UE's inactivity or detach. To detect UE's inactivity, the eNB uses an Inactivity Timer that restarts after a data packet transfer.

Figure 2 illustrates an example of the power consumption transitions the UE experiences while using both saving mechanisms. They are briefly described next:

- DRX: enables the UE to discontinuously receive Physical Downlink Control Channel (PDCCH). DRX is configured through DRX cycles. In each DRX cycle there are two phases. First, the UE monitors the PDCCH for a short period. Second, the UE stops monitoring the PDCCH for a long period. DRX saves UE's battery but still allows the network to reach the UE through Paging messages, or downlink control channels. There are short and long types of DRX cycles. In LTE, a UE can use both types. However, in NB-IoT, a UE in CE only uses long DRX cycles. This mechanism can be used while the UE is in RRC Connected or RRC Idle. For more information about DRX, see [5].
- PSM: allows the UE in RRC Idle state to enter deep sleep. In deep sleep, the UE is unreachable by the network, but it is still registered. A UE using PSM remains in deep sleep until a mobile originated transaction requires initiating a communication with the network. One example is the periodic Tracking Area Update (TAU) procedure (triggered by the expiration of an associated timer) or an uplink data transmission. Before entering PSM, the UE must be reachable by the network for a period of time. During this period, the UE may use DRX to enable possible downlink transmissions while saving power consumption. This period starts when the UE transits to RRC Idle state and its duration is controlled by the Active Timer (see Figure 2). For more information about PSM, see [6].

For NB-IoT, an extended DRX cycle of 10.24s is supported in RRC Connected. In RRC Idle, the maximum DRX cycle is 2.91 hours. For PSM, the maximum PSM time is 310 hours. The extension of both mechanisms implies a higher latency as the network will wait a longer period until it is able to reach the UE. However, it reduces power consumption of the UE.

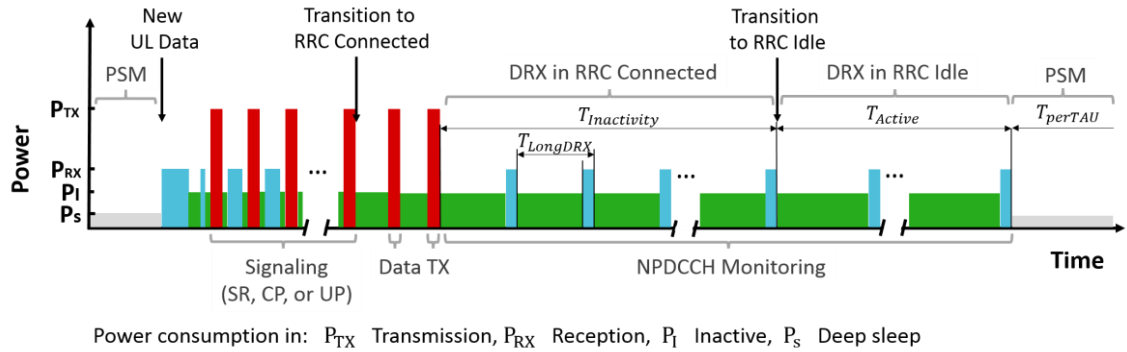

Figure 2: Example of power consumption transitions during a UE's connection.

## 3. Data transport in CIoT

In order to support mMTC communications, CIoT requires minimizing the signaling overhead, especially over the radio interface. In this section, we provide a description of the two CIoT EPS optimized procedures the 3GPP has introduced in Release 13 for that purpose: CP and UP [2]. Additionally, we describe the SR procedure as a reference. Figure 3 shows a joint view of the three procedures. Note that this diagram simplifies the particularities of each procedure. For detailed signaling flows, see [2].

During our description, we will differentiate between Mobile Originated (MO) and Mobile Terminated (MT) cases. In the MO case, a UE in RRC Idle state triggers the data transmission procedure because it has new traffic to send in uplink. In the MT case, it is the network that has new traffic to send in downlink to a UE in RRC Idle state.

### 3.1 Service Request (SR) procedure

In LTE, the transmission or reception of new data to/from a UE in RRC Idle state requires the establishment of an RRC connection. The procedure for that purpose is the SR. SR reestablishes the RRC connection between the UE and the eNB, and user plane bearers. After RRC connection establishment, the UE transits to RRC Connected state and the eNB is able to allocate radio resources. Consequently, the UE can send or receive data packets. After the data transmission, an S1 Release procedure is triggered to release the resources and the UE transits again to RRC Idle state. While the support of SR is optional for NB-IoT UEs, any NB-IoT UE that supports UP optimization shall also support SR.

For MO case, the UE in RRC Idle initiates the SR procedure (see Figure 3). As the UE

does not have an RRC connection active, it first needs to communicate with the network through a contention-based Random Access (RA). Then, the UE and eNB establish the RRC connection. Furthermore, to be able to send packets securely through the radio interface, the UE and eNB configure Access Stratum (AS) security. After successful setup of the AS security, the eNB reconfigures the RRC connection to finally establish a data radio bearer for the UE.

From this point on, the UE can send uplink data packets. Finally, the eNB and other core entities establish the rest of the user plane bearers to enable downlink data path. Later, if there are no user plane packets exchanged for a period, the eNB detects UE's inactivity and initiates the S1 Release procedure (see Figure 3).

For MT case, there are two possibilities to reach the UE in RRC Idle, depending on whether it is using DRX or PSM mechanism. If the UE is using DRX, it will listen to the network periodically. In this case, the network can send a Paging message to notify there is pending downlink traffic to deliver to the UE. After the UE recognizes the Paging message, it initiates the SR procedure as described for MO. However, if the UE is using PSM, it will be unreachable until the UE initiates either a MO transmission or the TAU procedure. In the latter case, the network benefits from the fact that the UE is in RRC Connected after performing the TAU to activate the SR procedure. In such case, after performing the TAU, the UE and the network perform the remaining steps to establish user plane bearers and AS security setup as for SR in MO case.

For downlink traffic, the performance of periodic TAUs to exit PSM may imply a higher latency than DRX. This is because of the period in deep sleep of PSM can be longer than the DRX cycle. However, PSM could extend the UE's battery lifetime if the traffic is sporadic, compared to DRX.

## 3.2 Control Plane CIoT EPS optimization (CP)

This optimization uses the control plane to forward the UE's data packets (see Figure 3). To do that, the data packets are sent encapsulated in Non Access Stratum (NAS) signaling messages to the MME (messages 5 and 6 of Figure 3). For NB-IoT UEs, the support of this procedure is mandatory.

Since CP uses control plane to forward data packets, the transmission or reception of messages is sent as NAS signaling messages between the UE and MME. Compared to conventional SR procedure, the UE avoids AS security setup and user plane bearers establishment required in each data transfer. Hence, it is more suitable for short data transactions.

When a UE transmits data in uplink, the NAS signaling message encapsulating the data packet can include a Release Assistance Information (RAI) field. This RAI field allows the UE to notify the MME if no further uplink or downlink data transmissions are expected, or only a single downlink data transmission subsequent to this uplink data transmission is expected. In such case, the MME can immediately trigger the S1 Release procedure (unless user plane bearers between eNB and Serving Gateway (SGW) are established). Hence, the RAI field enables the MME to reduce the period the UE is in DRX waiting for possible additional transmissions. Unfortunately, CP does not currently allow the application servers to notify the MME if no further data transmissions are expected.

Additionally, the UE or MME can trigger the establishment of the user plane bearers between eNB and SGW during data transmissions in CP. For example, if the size of the data transferred with CP exceeds a limit, the MME can initiate this procedure. This change of functionality implies the release of the specific CP user plane bearer between the MME and SGW, the establishment of the user plane bearers, and the setup of the AS security.

*3.3 User Plane CIoT EPS optimization (UP)*

The alternative data transmission procedure optimization is UP. It requires an initial RRC connection establishment that configures the radio bearers and the AS security context in the network and UE. After this, UP enables the RRC connection to be suspended and resumed by means of two new control procedures: Connection Suspend and Resume (see Figure 3). The support of this optimization is optional for NB-IoT UEs.

When the UE transits to RRC Idle state, the Connection Suspend procedure enables to retain UE's context at the UE, eNB, and MME. Later, when there is new traffic, the UE can resume the connection. To resume the RRC connection, the UE provides a Resume ID to be used by the eNB to access the stored context. By means of preserving the UE context instead of release it, the UE avoids AS security setup and RRC reconfiguration in each data transfer, compared to conventional SR procedure.

As the UP optimization utilizes the usual user plane connectivity, subsequent data packets can be transferred through the data paths. Therefore, UP is suitable for short or large data transactions.

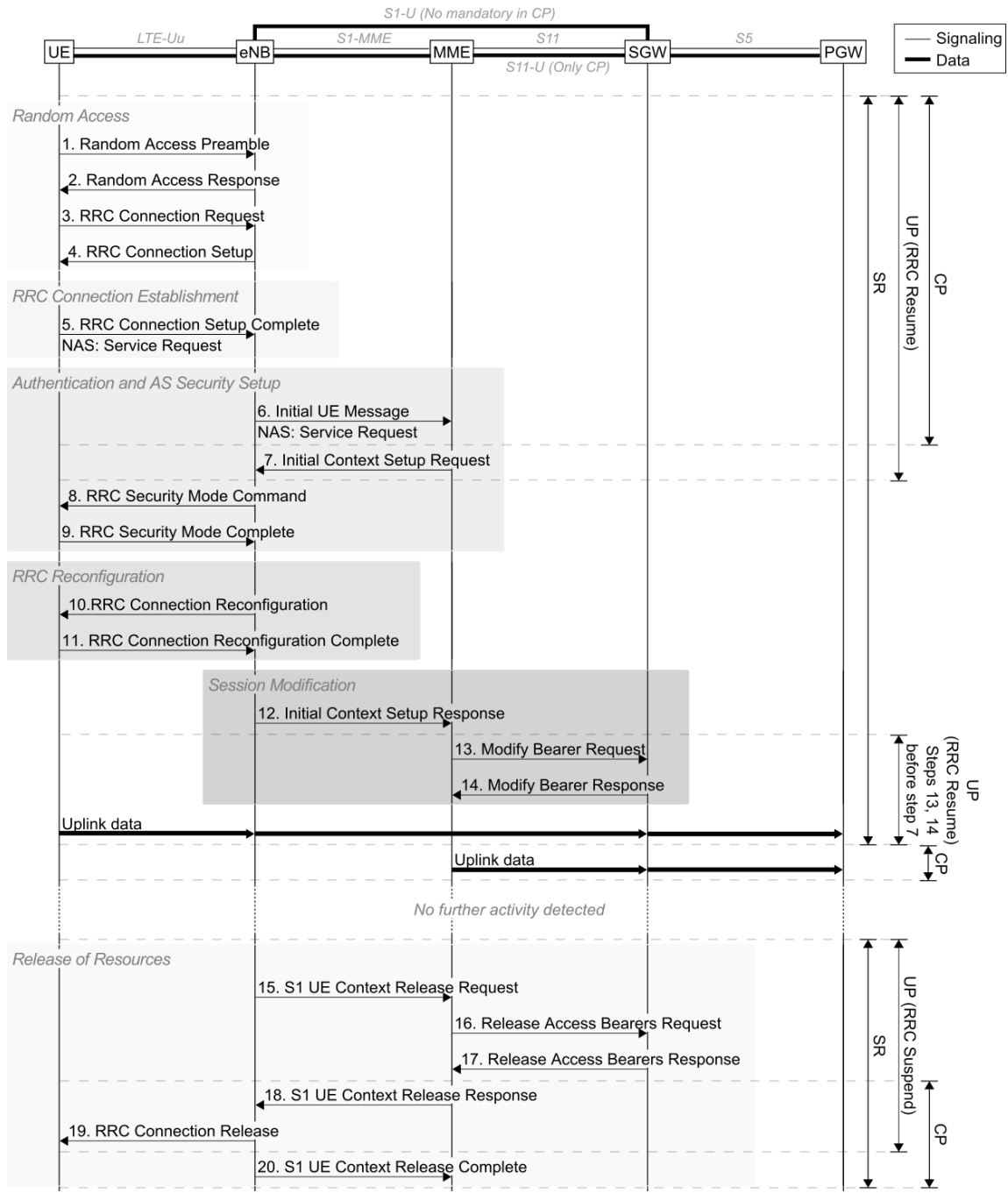

Figure 3: Summarized signaling diagram of MO data transport/RRC Resume and S1 Release/RRC suspend for SR, CP, and UP [2].

## 4. New functionalities to support data transmission optimizations

In order to support both CP and UP optimizations, LTE's network requires some functionality modifications. In this section, we list the most relevant modifications of each optimization. Note that network slicing may be an interesting solution to enable the introduction of these modifications and, therefore, facilitate the support of mMTC over a common network infrastructure [7].

*4.1 CP added network functionalities*

Generally, CP optimization relocates the user plane traffic. In LTE, this traffic is handled by the eNB and SGW. However, with CP the UE sends user plane data packets encapsulated within NAS messages via control plane to the MME. Hence, the MME acts as an intermediary between the eNB and SGW, and the CP optimization mainly implies the addition of user plane functionalities at the MME. Consequently, there is a significant impact on MME's functionalities and the conventional use of NAS security context, such as:

- Utilization of the S11 interface between MME and SGW to transfer user plane data through a new S11-U tunnel. This implies the addition of a user plane protocol stack in the MME. Particularly, the inclusion of the GPRS Tunnelling Protocol User plane (GTP-U) at this entity.
- A major increase of the processing load on the MME due to the processing of NAS data Protocol Data Units (PDUs). This implies prioritization and congestion handling between the NAS signaling PDUs and NAS data PDUs at the MME. Virtualized environments may address the support of this new load, as shown in [8]. Furthermore, the MME can supply the eNB with the UE's QoS profile to assist with resource prioritization decisions.
- Rate control of user data sent to and from a UE. This allows the network to protect its MME and the signaling radio bearers from the load generated by NAS data PDUs.
- Downlink user plane data buffering at the MME. CP enables SGW or MME buffering. Therefore, the MME may have to store data packets.
- The possibility of immediate release of radio resources triggered by the MME if the UE indicates the RAI in the NAS PDU. This only adds another cause for MME's initiated S1 Release procedure.

*4.2. UP added network functionalities*

Compared to CP, UP optimization involves lower impact in the network. The reason is UP utilizes conventional user plane to transfer data. However, some modifications are required to enable the UE and the network to resume the connection, such as:

- UE's context retained at UE, eNB, and MME.
- In order to support UE's mobility through different eNBs, there is a new control procedure defined, called Retrieve UE Context. This procedure allows inter eNB connection resumption when the UE resumes the connection in a new eNB.

**5. Performance**

In this section, we analyze the performance of the presented CIoT EPS optimizations assuming In-band NB-IoT deployment and three coverage levels: Extreme, Robust and Normal. The analytical model used is an extension of [9], which in turn also extends the model in [10]. The model in [9] has been expressly adapted to include specific NB-IoT features, such as radio resource allocation and channel configurations, power control formulas, and half-duplex transmission. Table 1 summarizes the main parameters' values considered in the evaluation.

| Energy consumption configuration ||
|---|---|
| Variable | Value |
| Deep sleep power consumption | 0.015 mW [11] |
| Inactive power consumption | 3 mW [11] |
| Reception power consumption | 90 mW [11] |
| Maximum Transmission power consumption | 545 mW [11] |
| Battery capacity | 5 Wh [1] |
| UE's transmit power for NPUSCH | Calculated as subclause 16.2.1.1.1 of [4], where: $M_{NPUSCH} = NumberOfSubcarriersOccupied$ $P_{O-NPUSCH,c} = -100 dBm$ $\alpha_c = 1$ |
| UE's transmit power for NPRACH | Calculated as subclause 16.3.1 of [4], where: $InitialReceivedTargetPower = -100 dBm$ $\Delta_{preamble} = 0$ $PowerRampingStep = 0 dB$ |
| DRX in Connected state (UE's Inactivity Timer) | UP, SR: 0s<br>CP: 5 NPDCCH periods |
| DRX in Idle state (UE's Active Timer) | UP, SR: 10s + 2 DRX cycles [5]<br>CP: 0s (UL, UL-ACK and DL-ACK cases)<br>CP: 10s + 2 DRX cycles (DL case) |
| DRX long cycle | 2.048s + 1 NPDCCH period |
| Preamble detection probability | Preamble: $1 - e^{-i}$, where i indicates the i-th preamble transmission [12]<br>Other packets: 1 |
| UE's traffic model | Data packets: 20B payload + 44B overhead [1]<br>Data acknowledgments: 0B payload + 44B overhead [1] |

| NB-IoT design | | | |
|---|---|---|---|
| Radio conditions [13][14][15] | Normal | Robust | Extreme |
| *Target Maximum Coupling Loss (dB)* | ≃144 | ≃154 | ≃161 |
| *Subcarrier Spacing (kHz)* | 15 | 15 | 3.75 |
| *Number of subcarriers in uplink for a burst* | 12 | 3 | 1 |

| | Modulation | QPSK | QPSK | BPSK |
|---|---|---|---|---|
| | Modulation and Coding Scheme (MCS) | 9 | 3 | 0 |
| | Narrowband Physical Downlink Control Channel (NPDCCH) repetitions | 1 | 64 | 512 |
| | Narrowband Physical Downlink Shared Channel (NPDSCH) repetitions | 1 | 32 | 256 |
| | Narrowband Physical Uplink Shared Channel (NPUSCH) repetitions | 2 | 16 | 1 |
| | Narrowband Physical Random Access Channel (NPRACH) repetitions | 1 | 8 | 32 |
| Ratio of the total resources for each radio condition | 33% | | | |
| NPDCCH design | Format: 0<br>Aggregation Level: 2<br>Periodicity: $T_{NPDCCH} = R_{MAX} \cdot G$ [4], where for each CL:<br>Normal: $R_{MAX} = 1$ and $G = 32$<br>Robust: $R_{MAX} = 64$ and $G = 1.5$<br>Extreme: $R_{MAX} = 512$ and $G = 1.5$ | | | |
| Timing relationships [4] | | | | |
| *Start of NPUSCH transmission after the end of its associated NPDCCH* | 8ms | | | |
| *Start of NPDSCH transmission after the end of its associated NPDCCH* | 4ms | | | |
| Random Access Opportunity | 40 ms [15] | | | |

Table 1: Main parameters used in performance evaluation.

The evaluation of the optimizations is done for a small data transmission of one data packet in the following cases:

- Uplink (UL): The NB-IoT UE sends a report to the IoT application server.
- Uplink with an acknowledgment (UL-ACK): The same as UL case but the server replies with a downlink acknowledgment packet as a confirmation.
- Downlink (DL): The NB-IoT UE receives an application layer command from the application server.
- Downlink with an acknowledgment (DL-ACK): The same as DL case but the UE replies with an uplink report.

For the evaluation, we assume the NB-IoT UE employs PSM to reduce battery consumption. We suppose that the pattern of power consumption transitions is as depicted in Figure 2. For UL and UL-ACK cases, we assume the NB-IoT UE performs a periodic TAU procedure with a period of 5 days. For DL and DL-ACK cases, the periodic TAU is configured with the same frequency as the downlink traffic.

Figure 4 shows the battery lifetime of an NB-IoT UE for UL case. The figure presents the results of the three coverage level considered and different Inter Arrival Times (IATs). The evaluation includes CP, UP, SR, and a baseline consumption due to PSM. The energy consumption analysis does not include UE processing consumption. The figure includes results from [11], which uses a configuration similar to our Normal coverage. This evaluation estimates the energy consumption of an uplink transmission after a RA but without the specific signaling required to perform the data transmission procedure. These results are included to confirm that our model obtains similar battery lifetime as [11] under similar configuration.

The battery lifetime decreases significantly for short IATs. In Normal coverage, for IATs shorter than 5-10 hours the consumption caused by the synchronization time in the initial RA procedure and the inactive time spent in DRX make up a large proportion of the overall consumption (58% in UL case for UP with an IAT of 1h). For IATs longer than 5-10 hours, the overall energy consumption is dominated by the time spent in PSM (84% in UL case for UP with an IAT of 10h).

The total period the NB-IoT UE spends performing DRX is comprised by Inactivity and Active timers (see Table 1). Keeping the UE in DRX decreases the battery lifetime. However, it reduces the probability of reestablishing the connection once it has been released, which may be interesting for traffic comprised by bursts of packets not considered in this study. CP enables the MME to know if there is more pending traffic through the RAI notification. Then, the UE avoids the need of DRX. For Normal coverage, this improves the battery lifetime up to 78% compared to UP. However, if the required repetitions increase, there is no noticeable impact. If UP is configured with the same Inactivity and Active timers as CP, both optimizations provide similar results.

When the UE has poor radio conditions and changes to a worse coverage level, there is a significant reduction of battery lifetime. This is mainly due to the repetitions required and the lower MCS used. For worse coverage levels, the energy consumption is dominated by the messages exchanged from the completion of the RA procedure up to the end of the data transmission for almost the entire range of IATs (35% and 49% in UL case for UP with an IAT of 10h for Robust and Extreme coverages, respectively). The rest of the energy consumption is due to PSM. However, for very long IATs, the energy consumption in PSM prevails (67% and 42% in UL case for UP with an IAT of 24h for Robust and Extreme coverages, respectively). For Extreme coverage, the considered subcarrier spacing of 3.75kHz partly mitigates the battery lifetime reduction due to the reduced number of NPUSCH repetitions (see Table 1).

For the other evaluated cases (UL-ACK, DL, and DL-ACK), we obtain similar results for all procedures in Normal coverage. Except for CP in DL case, where the MME cannot receive the RAI notification in the downlink NAS PDU. Hence, CP optimization requires the DRX mechanism as UP, and therefore the results of both procedures become similar. In Robust and Extreme coverages, UL-ACK results are similar to UL case. However, for DL and DL-ACK there is a significant battery lifetime reduction compared to UL case (up to 30% in DL, and 55% in DL-ACK). This is due to downlink traffic being handled by means of periodic TAUs, that implies the transmission/reception of heavy signaling packets.

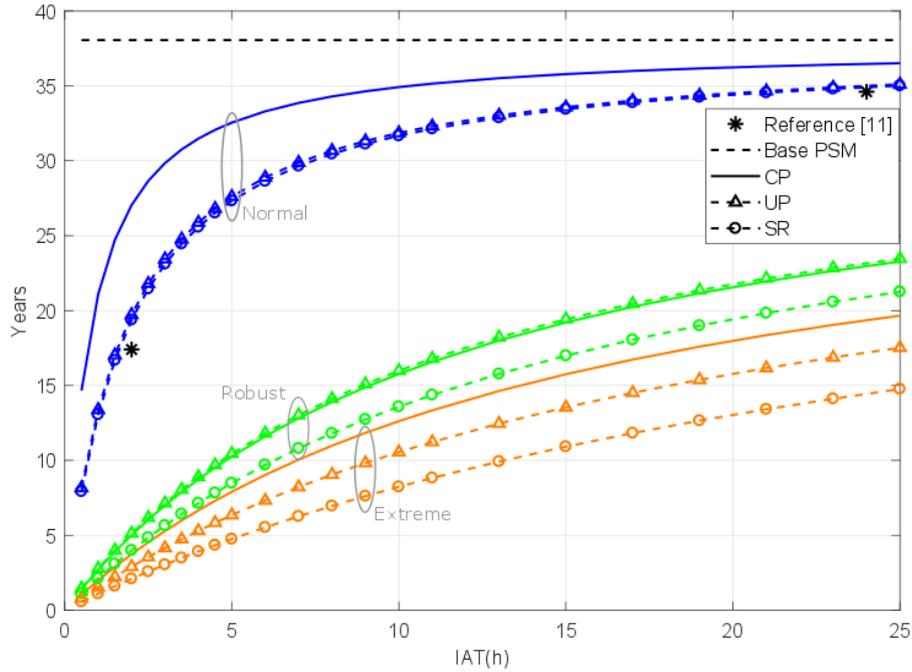

Figure 4: NB-IoT UE battery lifetime considering UL case.

Figure 5 shows CP and UP capacity gain relative to SR procedure for different coverage levels and all cases considered in this work, assuming an NB-IoT UEs' IAT of 1 hour. From all cases evaluated, UL shows the best results, as CP and UP achieve the greatest reduction in signaling compared to SR. In this case, the capacity gain relative to SR of both procedures reach 162% and 120% in Normal coverage for CP and UP, respectively. The signaling inefficiency of SR was also shown in [10] where the SR procedure was compared to an assumed lightweight-signaling access for an LTE system.

Our results show that the radio channel limiting the capacity varies depending on the considered case and coverage level. Furthermore, the use of the radio channels is different for each procedure. For Normal coverage and UL case, the capacity is mostly limited by the uplink channels' resources. On the contrary, for UL case and the rest of coverage levels, the capacity limitation comes from the downlink. This limitation is due to the number of repetitions in both NPDCCH and NPDSCH, legacy PDCCH reserved resources of In-band's deployment, and the sharing of downlink subframes with downlink signals. Moreover, CP reduces the required resources at NPDCCH compared to UP and SR. Therefore, for worst coverage levels increasingly limited by NPDCCH resources, its gain keeps increasing too.

Regarding UL and DL evaluation, DL gains reach something less than 2 times the gains of UL. This is due to downlink traffic being handled by means of periodic TAUs. Therefore, causing additional signaling for the network in both uplink and downlink channels. Particularly in DL and DL-ACK cases, CP achieves slightly worse results than UP. This is because CP increases the load at NPDSCH, compared to UP, and both cases are more intensive in NPDSCH.

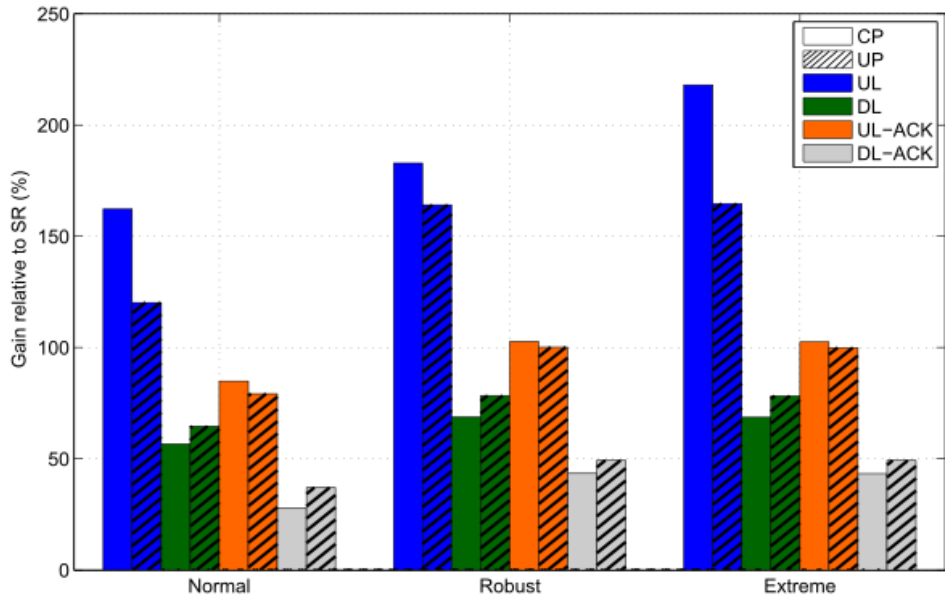

Figure 5: CP and UP capacity gain relative to SR in different coverage levels and cases.

Additionally, for larger data payloads than the 20B evaluated, there would be a significant UE's battery lifetime reduction if the number of NPUSCH repetitions is high. In terms of capacity, CP and UP would achieve a lower capacity gain relative to SR in cases limited by uplink channels. This is due to the signaling overhead could be negligible compared to the data transferred.

## 6. Conclusions

In this article, we provide an overview and a performance analysis of the new small data transmission optimizations included in Release 13 for CIoT. The optimizations analyzed are: Control Plane CIoT EPS (CP) and User Plane CIoT EPS (UP). While UP uses the conventional user plane to transfer data packets, CP uses the control plane, implying new functionalities and an increased load at the MME.

We analyze a small data transfer in NB-IoT. For this purpose, we evaluate the device's battery lifetime and the cell capacity gain relative to SR of both optimizations, for three different coverage levels. Regarding battery lifetime, in Normal coverage for IATs larger than 5-10 hours the overall energy consumption is dominated by the time spent in PSM. On the contrary, for short IATs, the consumption caused by the data transmission procedure prevail. For Robust and Extreme coverages, the energy consumption is dominated by the messages exchanged after RA for almost the entire range of IATs. However, for very long IATs PSM still dominates the consumption.

Regarding the cell capacity evaluation, the results highlight both optimizations reach considerably capacity gain relative to SR. In UL case, CP and UP achieve gains of 162% and 120% in Normal coverage, respectively. If PSM is used to extend battery lifetime, DL gains reach something less than 2 times the gains of UL. This is due to the additional signaling generated to perform the periodic TAU to manage downlink traffic.

The comparison of CP and UP optimizations yields similar results, except for some specific configurations. CP achieves up to 78% of battery lifetime in UL case and Normal coverage due to its RAI indication. Furthermore, the use of less resources at the NPDCCH improves CP's cell capacity gain for UL case. However, CP is not convenient for long data transmissions, as the network is expected to force the UE to establish the data bearers if a maximum number of messages is exceeded.

Regarding future work, several challenges lie ahead for NB-IoT. 3GPP's Release 13 has specified the baseline NB-IoT. Beyond Release 13, Release 14 includes support for positioning, multi-cast and non-anchor carrier operation for NB-IoT. Furthermore, Release 15 work items include Time-Division Duplexing support and enhancements to reduce latency.

**Acknowledgments**

This work is partially supported by the Spanish Ministry of Economy and Competitiveness and the European Regional Development Fund (Projects TIN2013-46223-P, and TEC2016-76795-C6-4-R), and the Spanish Ministry of Education, Culture and Sport (FPU Grant 13/04833).

**Biographies**

**Pilar Andres-Maldonado** (pilaram@ugr.es) received the M.Sc. degree in telecommunications engineering from the University of Granada, Spain, in 2015. She is currently a Ph.D. candidate in the Department of Signal Theory, Telematics and Communications of the University of Granada. Her research interests include Machine-to-Machine communications, NB-IoT, 5G, LTE, virtualization, and software-defined networks.

**Pablo Ameigeiras** (pameigeiras@ugr.es) received his M.Sc.E.E. degree in 1999 from the University of Malaga, Spain. He performed his Master's thesis at the Chair of Communication Networks, Aachen University, Germany. In 2000 he joined Aalborg University, Denmark, where he carried out his Ph.D. thesis. In 2006 he joined the University of Granada, where he has been leading several projects in the field of LTE and LTE-Advanced systems. Currently his research interests include 5G and IoT technologies.

**Jonathan Prados-Garzon** (jpg@ugr.es) received the M.Sc. degrees in telecommunications engineering and multimedia systems from the University of Granada, Granada, Spain, in 2011 and 2012, respectively. He is currently a Ph.D. candidate in the Department of Signal Theory, Telematics and Communications, University of Granada. He was granted a Ph.D. fellowship by the Education Spanish Ministry in September 2014. His research interests include 5G mobile network architectures, 3GPP LTE systems, network functions virtualization, software-defined networks, and machine-to-machine communications.

**Jorge Navarro-Ortiz** (jorgenavarro@ugr.es) is Associate Professor at the Department of Signal Theory, Telematics and Communications of the University of Granada. He obtained his M.Sc. in Telecommunications Engineering at the University of Malaga in 2001. Afterwards, he worked at Nokia Networks, Optimi/Ericsson and Siemens. He started working as Assistant Professor at the University of Granada in 2006, where he got his Ph.D. His research interests include wireless technologies for IoT such as LoRaWAN and 5G among others.

**Juan M. Lopez-Soler** (juanma@ugr.es) is professor at the Department of Signals, Telematics and Communications (University of Granada). In 1991-92 he joined the Institute for Systems Research at the University of Maryland. He is the head of the WiMuNet Lab at the University of Granada. He has participated in 24 research projects, has advised five Ph.D. theses, and has published 24 papers and more than 40 workshops/conferences. His research interests include middleware, multimedia communications, and 5G networking.